\documentclass{article} 

\usepackage{iclr2027_conference,times}

\usepackage{wrapfig}
\usepackage{booktabs}
\usepackage{caption}
\usepackage{tabularray}
\UseTblrLibrary{booktabs}
\usepackage{url}
\usepackage{multirow}
\usepackage{graphicx}
\usepackage[table]{xcolor}
\usepackage{float}
\usepackage{enumitem}
\usepackage{tabularx}
\usepackage{amsmath,amssymb}
\usepackage[most]{tcolorbox}
\usepackage{siunitx}

\newtcolorbox{promptbox}[1][]{
  enhanced,
  breakable,
  colback=gray!5,
  colframe=black!40,
  boxrule=0.6pt,
  arc=2pt,
  left=6pt,
  right=6pt,
  top=6pt,
  bottom=6pt,
  #1
}

\usepackage{hyperref}
\hypersetup{
  colorlinks=true,
  linkcolor=blue,
  filecolor=magenta,
  urlcolor=cyan,
}

\title{FATE: Frame-level Audio-Visual Temporal Embedding}

\author{
\textbf{Kaisi Guan}\textsuperscript{1} \quad
\textbf{Bingzi Zhang}\textsuperscript{1} \quad
\textbf{Xihua Wang}\textsuperscript{1} \quad
\textbf{Ying Ba}\textsuperscript{1} \quad
\textbf{Xin Cheng}\textsuperscript{1} \\
\textbf{Yijing Chen}\textsuperscript{1} \quad
\textbf{Ruihua Song}\textsuperscript{1}\thanks{Corresponding author.} 
\\
\normalfont
\textsuperscript{1}\,Gaoling School of Artificial Intelligence,
Renmin University of China \\
\texttt{guankaisi@ruc.edu.cn}
}

\iclrfinalcopy

\begin{document}

\pagestyle{plain}

\maketitle

\thispagestyle{plain}
\pagestyle{plain}

\begin{abstract}
    When a dog opens its mouth and barks, humans naturally recognize what the sound is and when it occurs. Building audio-visual models with this same ability requires representations that capture both semantic and temporal alignment. Current approaches fall short on one side or the other: embedding models capture semantics but lose temporal information; synchronization models capture temporal offsets but lack semantic understanding.
    To bridge this gap, we propose \textbf{FATE}, \textbf{F}rame-level \textbf{A}udio-visual \textbf{T}emporal \textbf{E}mbedding. Unlike prior embedding models that pool each modality into a single embedding and discard temporal information, FATE retains frame-level sequences, aligns them on the physical timeline, and computes similarity over strictly aligned frame pairs. Unlike synchronization models that output only an offset prediction, FATE encodes synchronization in a reusable embedding space, trained with a joint objective combining cross-video semantic and within-video temporal contrastive learning to capture both \emph{what} sounds and \emph{when} it occurs. 
    Across three tasks, FATE surpasses the strongest baseline on temporal and semantic retrieval by a large margin, matches fully supervised methods on event localization in a zero-shot setting, and achieves the best correlation with human judgments as a generation evaluation metric. The source code can be found at \texttt{https://github.com/guankaisi/FATE}.
\end{abstract}

\section{Introduction}
\label{sec:intro}
When a dog opens its mouth to bark, the barking sound occurs at precisely the same moment. Any semantic contradiction or temporal misalignment immediately feels unnatural to human observers. This tight coupling between \emph{what} and \emph{when} is the foundation of human audio-visual perception.

Similarly, models designed to emulate human perception, like multimodal understanding or generation, must also capture both semantic and temporal information. Existing approaches, however, typically emphasize one dimension at the expense of the other.
\begin{figure}[h!]
    \centering
    \includegraphics[width=\textwidth]{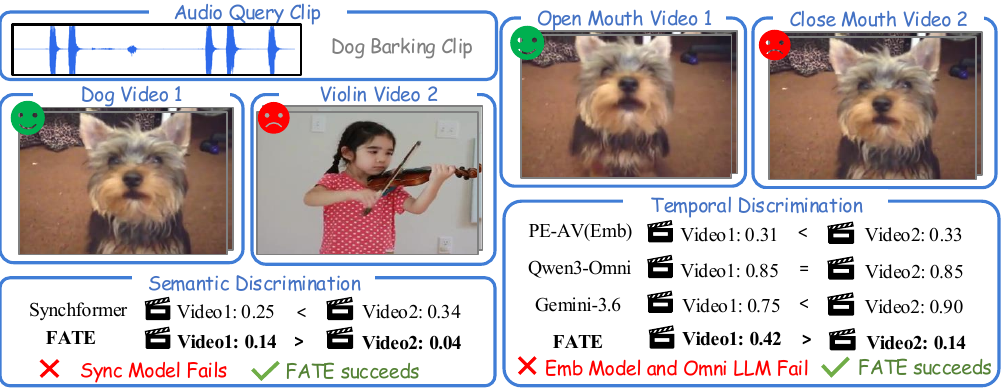}
   \caption{Semantic (what) vs. temporal (when) discrimination for a dog-barking audio query. Top: the synchronization model scores the unrelated violin video higher than the barking dog. Bottom: the embedding model (PE-AV) and Omni LLMs score the unsynchronized segment (mouth closed) higher than or equal to the synchronized one, whereas FATE succeeds in both dimensions. Embedding and synchronization scores are normalized to $[0,1]$ (higher is better; Synchformer's offset output is inverted); Omni LLMs rate each clip independently via prompting
(details in Appendix~\ref{app:mllm_prompt}).}
    \label{fig:teaser}
    \vspace{-3mm}
\end{figure}
Audio-visual embedding models~\citep{radford2021clip,DBLP:conf/icassp/WuCZHBD23,assran2025vjepa2,tuncay2025audiojepa,guzhov2021audioclip,girdhar2023imagebind,araujo2025cav-mae-sync,vyas2025pe-av} train dual encoders with contrastive objectives to project audio and video into a shared embedding space. Although effective at semantic matching, they often discard fine-grained temporal information: as shown in Fig.~\ref{fig:teaser} (bottom), such models score an unsynchronized segment even higher than the synchronized one. Conversely, synchronization models~\citep{chen2021vggsync,iashin2024synchformer} capture temporal offsets, but their task-specific objectives yield weak semantic representations: in Fig.~\ref{fig:teaser} (top), the synchronization model fails to distinguish the barking dog from unrelated violin. Nor do recent Omni LLMs~\citep{Qwen3-Omni,geminiteam2023gemini,chen2025chronusomniimprovingtimeawareness} close this gap: their low sampling rates (e.g., 2\,fps for Qwen3-Omni) preclude fine-grained synchronization, and they produce no reusable embeddings for downstream tasks(Fig.~\ref{fig:teaser}, bottom). Consequently, a unified embedding representation that jointly captures \emph{what} sounds and \emph{when} it occurs remains lacking.

In this paper, we propose \textbf{FATE}, \textbf{F}rame-level \textbf{A}udio-visual \textbf{T}emporal \textbf{E}mbedding. The key idea is simple: instead of compressing each modality into a single global vector, FATE preserves frame-level features and aligns the two modalities on the physical time axis, so that cross-modal similarity naturally reflects both semantic and temporal alignment. We train FATE with a joint objective combining semantic contrastive learning across videos and temporal soft-contrastive learning within each video, encouraging the embedding space to capture both \emph{what} sounds and \emph{when} they occur.

We evaluate FATE on three downstream tasks. For temporal cross-modal retrieval, FATE outperforms the strongest synchronization baseline by over 13 points in R@3, while simultaneously achieving over 20 points higher semantic retrieval accuracy. For audio-visual event localization, FATE in a zero-shot setting, surpasses supervised methods with 48.3\% average accuracy. As an automatic metric for audio-visual generation, FATE agrees most closely with human synchronization judgments among all compared metrics.
In summary, our main contributions are as follows:




\noindent (1) We introduce \textbf{FATE}, a reusable frame-level audio-visual embedding model. It preserves temporal feature sequences, aligns audio and video in physical time, and scores time-aligned frame pairs, representing synchronization without task-specific prediction heads.

\noindent (2) We train FATE with cross-video semantic contrastive learning and within-video temporal soft-contrastive learning. Controlled ablations show that the two objectives are complementary and that offset-aware soft targets improve temporal discrimination.

\noindent (3) We demonstrate FATE's transfer across temporal cross-modal retrieval, zero-shot event localization, and synchronization evaluation of generated videos. Without task-specific fine-tuning, FATE leads the retrieval baselines, matches supervised localization methods, and achieves the strongest agreement with human judgments among the compared metrics.
\section{Related Work}
\subsection{Audio-Visual Representation Learning}
Multi-modal representation learning has advanced significantly, driven by the success of contrastive learning~\citep{oord2018infonce}. CLIP~\citep{radford2021clip} pioneers this direction by aligning image and text modalities. This paradigm naturally extends to the audio-visual domain: AudioCLIP~\citep{guzhov2021audioclip}, ImageBind~\citep{girdhar2023imagebind} LanguageBind~\citep{zhu2023languagebind} and PE-AV~\citep{vyas2025pe-av} scale this idea further to audio, image, text and video. However, all of these approaches capture \textit{what} co-occurs across modalities but not \textit{when}. Recent omni-modal LLMs~\citep{Qwen3-Omni,geminiteam2023gemini,chen2025chronusomniimprovingtimeawareness} also process audio and video jointly, but their sparse visual sampling (e.g., 2\,fps in Qwen3-Omni) is far coarser than the granularity that synchronization requires, and they yield no reusable embeddings for downstream task; Fig.~\ref{fig:teaser} illustrates this failure. This gap motivates our frame-level design.

\subsection{Audio-Visual Synchronization Models}
Audio-visual synchronization aims to determine whether audio and video clips are temporally coherent.
SyncNet~\citep{chung2017syncnet} pioneered this direction with two-tower CNNs for lip-speech synchronization, and AVST~\citep{chen2021vggsync} generalized it to open-domain videos. SparseSelector~\citep{iashin2022sparseselector} and Synchformer~\citep{iashin2024synchformer} further moved from binary detection to quantitative offset estimation. On another front, CAVP~\citep{luo2023cavp} grounds synchronization in temporal contrastive learning, while PEAVS~\citep{goncalves2024peavs} directly regresses human opinion scores.
These models output a single prediction (an offset or a score) but do not yield reusable representations, making them difficult to transfer to other tasks. In contrast, FATE learns a general-purpose embedding space where synchronization is reflected in frame-level embeddings, making it applicable across diverse downstream scenarios.

\subsection{Evaluating Audio-Visual Generative Models}
Audio-visual generative models~\citep{wang2025jointdit,guan2025bridgedit,liu2025javisdit,low2025ovi,hacohen2026ltx2,seedance2025seedance,tiva,lova,vaflow,cheng2026vssflow} are advancing rapidly, yet reliable evaluation of synchronization remains limited~\citep{huang2023vbench,zheng2025vbench2,huang2025vbench++,guan2025etva,cheng2026syncdpoenhancingtemporalsynchronization}. Existing metrics fall into three paradigms. Rule-based metrics such as AV-Align~\citep{yariv2023avalign} match optical flow against audio onsets, but these low-level cues break down on complex generative artifacts. Offset predictors such as DeSync~\citep{iashin2024synchformer} regress a global temporal shift, yet synchronization errors in generated videos are rarely uniform offsets. Embedding-based scores such as CAVP~\citep{luo2023cavp} measure similarity in the
embedding space, but pool away the temporal dimension that synchronization depends on. FATE addresses all three by scoring synchronization over frame-level embeddings.
\begin{figure*}[t]
    \centering
    \includegraphics[width=\textwidth]{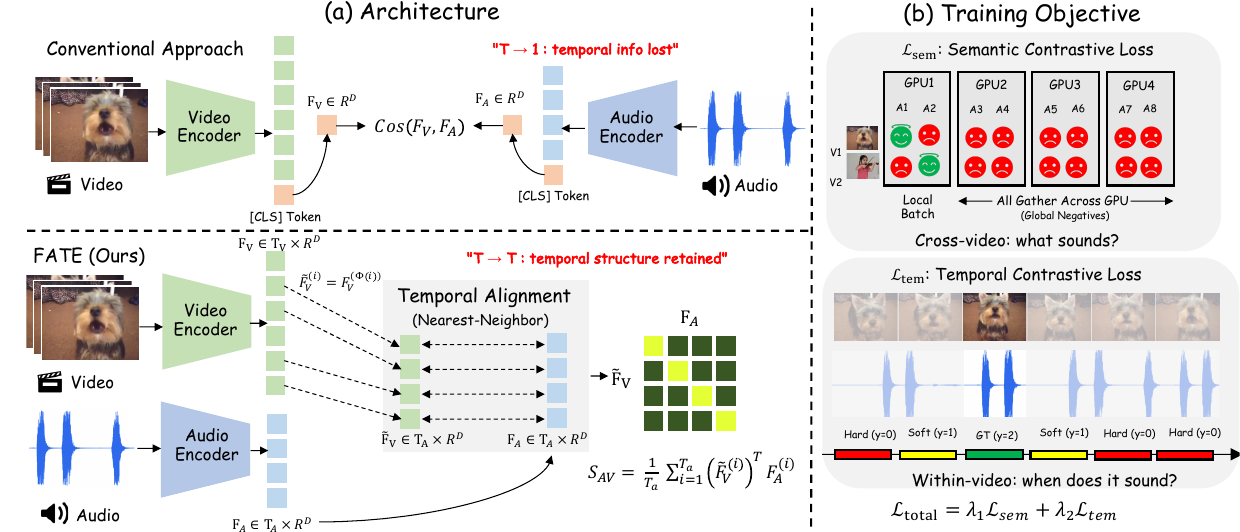}
    \caption{\textbf{Overview of FATE.} (a) \textbf{Architecture.} Conventional models pool encoder outputs into a single vector per modality ($T\!\to\!1$), discarding temporal structure. FATE retains frame-level tokens ($T\!\to\!T$), aligns audio and video by nearest-neighbor resampling, and averages similarities over time-aligned pairs. (b) \textbf{Training objective.} The semantic loss $\mathcal{L}_{\mathrm{sem}}$ separates videos using cross-GPU negatives, whereas the temporal loss $\mathcal{L}_{\mathrm{temp}}$ distinguishes offsets within each video using offset-dependent soft targets (Eq.~\ref{eq:soft_label}).}
    \label{fig:main}
    \vspace{-5mm}
\end{figure*}

\section{Method}
\label{sec:method_overview}
 
\subsection{Problem Formulation}
We aim to learn audio-visual representations that jointly encode semantic and temporal alignment. Let $V \in \mathbb{R}^{T \times H \times W \times 3}$ denote a video clip of $T$ frames, and let $A \in \mathbb{R}^{L \times C}$ denote its audio representation with $L$ temporal frames and $C$ channels. A dual-encoder architecture maps the two modalities into a shared embedding space:
\begin{equation}
    F_V = E_v(V), \quad F_A = E_a(A),
\end{equation} 
and the cross-modal similarity is computed as $S_{AV} = \mathrm{Sim}(F_A, F_V)$. 
We require $S_{AV}$ to satisfy two properties: (i) \emph{semantic discrimination}: $S_{AV}(V, A) > S_{AV}(V, A')$ for audio $A'$ drawn from a semantically different clip; and (ii) \emph{temporal discrimination}: $S_{AV}(V, A) > S_{AV}(V, A_{\Delta t})$ for any temporally shifted version $A_{\Delta t}$ of the same audio~($\Delta t \neq 0$). Both properties are required symmetrically in the video direction, i.e., $S_{AV}(V, A) > S_{AV}(V', A)$ and $S_{AV}(V, A) > S_{AV}(V_{\Delta t}, A)$.
Conventional embedding models satisfy (i) but not (ii). Synchronization models, which train without cross-video contrast, satisfy (ii) but fail to separate semantically different clips, violating (i). FATE satisfies both: by retaining frame-level embedding and defining $\mathrm{Sim}$ over temporally aligned frame pairs, semantic and temporal discrimination are encoded in one single similarity.
 
\subsection{Architecture of FATE}
\label{sec:method_arch}
Fig.~\ref{fig:main} illustrates the architecture of FATE. We build on the Perception Encoder Audio-Visual (PE-AV)~\citep{vyas2025pe-av}, retaining its dual-encoder backbone but discarding the global pooling head. The resulting pipeline operates in three stages: frame-level feature extraction, temporal alignment, and frame-level similarity computation.
 
\noindent\textbf{Frame-level feature extraction.}
FATE bypasses the pooling step and retains full tokens. Given video $V$ and audio $A$, the dual encoders produce $F_V \in \mathbb{R}^{T_v \times D}$ and $F_A \in \mathbb{R}^{T_a \times D}$, where $D$ is the shared embedding dimension and all frame features are $\ell_2$-normalized. The token counts $T_v$ and $T_a$ are determined by the video frame rate and the audio spectral hop length; in practice video encoders produce more tokens per second, yielding $T_v > T_a$ for the same clip duration. Element-wise similarity requires sequences of identical length, which motivates the alignment step below.
 
\noindent\textbf{Temporal alignment.}
The two encoders sample the timeline independently and share only one common coordinate system: physical time. Comparing the $i$-th video token with the $i$-th audio token without alignment is therefore meaningless, as they cover different time intervals. We align the sequences with two design choices, both grounded in the asymmetry between the modalities. First, we use audio as the \emph{temporal anchor} and downsample video to length $T_a$: video is temporally redundant --- adjacent frames are near-identical, so subsampling loses little --- whereas audio carries transient synchronization cues that resampling would destroy. Second, we use \emph{nearest-neighbor} rather than linear interpolation, so that every aligned token remains an original frame feature; linear blending would fabricate intermediate states with no physical counterpart. Both choices are validated against their alternatives (linear interpolation, repeat padding, and upsampling) in Table~\ref{tab:ablation_design} (ii). Formally, for each audio time step $i \in \{1, \dots, T_a\}$, the aligned visual feature is:
\begin{equation}
    \tilde{F}_V^{(i)} = F_V^{(\phi(i))}, \quad
    \phi(i) = \min\!\left(\mathrm{round}\!\left(i \cdot
    \tfrac{T_v}{T_a}\right),\, T_v\right).
\end{equation}
yielding an aligned visual sequence $\tilde{F}_V \in \mathbb{R}^{T_a \times D}$.
 
\noindent\textbf{Frame-level similarity.}
With both modalities on the same timeline, the cross-modal similarity is the
mean inner product over aligned frame pairs:
\begin{equation}
    S_{AV} = \frac{1}{T_a} \sum_{i=1}^{T_a}
    \left(\tilde{F}_V^{(i)}\right)^{\!\top} F_A^{(i)}.
    \label{eq:frame_level}
\end{equation}
This traces the main diagonal of the $T_a \times T_a$ pairwise
similarity matrix, enforcing temporal order by construction. Under a temporal
shift of $\Delta t$, the true correspondences move to the $\Delta t$-th
off-diagonal, so $S_{AV}$ decays as the fraction of correctly aligned content
shrinks --- an effect empirically visible as the sharp unimodal peak at
$\Delta t = 0$ in Fig.~\ref{fig:sensitivity}. This built-in penalty is what
FATE exploits for synchronization-aware retrieval and evaluation, without any
task-specific prediction head.
 
\subsection{Training Strategy}
\label{sec:training}
 
FATE is trained with two complementary objectives: semantic contrastive loss $\mathcal{L}_{sem}$ that separates embeddings \emph{across} videos, and temporal soft-contrastive loss $\mathcal{L}_{temp}$ that supervises fine-grained alignment \emph{within} each video. The total objective is their weighted sum, which together captures both \textit{what} sounds and \textit{when} it occurs.
\begin{equation}
    \mathcal{L}_{total} = \lambda_1 \mathcal{L}_{sem}
    + \lambda_2 \mathcal{L}_{temp},
\end{equation}
We detail each term below:
 
\noindent\textbf{Semantic contrastive loss.}
We apply symmetric InfoNCE over a batch of $B$ audio-visual pairs. For each video anchor $v_i$, its paired audio $a_i$ is the positive and all other
in-batch audios are negatives:
\begin{equation}
    \mathcal{L}_{v2a} = -\frac{1}{B} \sum_{i=1}^{B}
    \log \frac{\exp(S_{AV}(v_i, a_i) / \tau)}
    {\sum_{j=1}^{B} \exp(S_{AV}(v_i, a_j) / \tau)},
\end{equation}
with $\mathcal{L}_{a2v}$ computed symmetrically and
$\mathcal{L}_{sem} = \frac{1}{2}(\mathcal{L}_{v2a} + \mathcal{L}_{a2v})$.
Since frame-level embeddings limit the per-GPU batch size, we all-gather embeddings across $G$ GPUs, expanding the negative pool from $B$ to $B \times G$ at no extra per-GPU memory cost (Fig.~\ref{fig:main}(b)). 
 
\noindent\textbf{Temporal soft-contrastive loss.}
The semantic loss structures the embedding space \emph{across} videos but provides no supervision \emph{within} a video: it cannot distinguish a synchronized clip from a slightly shifted one. The temporal loss fills this gap by casting alignment as probability matching over candidate offsets. Given a 2-second video anchor $v$, we slide a 2-second window over its 10-second source video with a 0.5-second stride, yielding $K = 17$ candidate audio clips whose offsets $\Delta t_k$ range from fully aligned to strongly misaligned.
The supervision target can be any non-increasing function of the absolute offset; we instantiate it as a raised-cosine kernel,
\begin{equation}
    y(\Delta t_k) =
    \begin{cases}
        \cos^2\!\left(\dfrac{\pi \Delta t_k}{2T_0}\right)
            & |\Delta t_k| \le T_0, \\[6pt]
        0   & \text{otherwise},
    \end{cases}
    \label{eq:soft_label}
\end{equation}
which decays smoothly from its peak at $\Delta t_k = 0$ to zero at the support boundary $T_0$. We set $T_0 = 1$\,s so that the half-maximum point, $0.5$\,s, matches the psychophysical threshold below which humans can barely perceive audio-visual asynchrony~\citep{Vatakis2006AudiovisualSP}. The smooth roll-off also matches the structure of the candidates: adjacent clips share 75\% temporal overlap, so assigning them zero weight would force the model to separate similar content. For example, on our 0.5s grid, Eq.~\eqref{eq:soft_label} yields relative weights $2\!:\!1\!:\!0$ for the aligned, adjacent, and distant candidates; the kernel width is further validated by ablation (Table~\ref{tab:ablation_design}(i)). Normalizing $p_k = y(\Delta t_k) / \sum_m y(\Delta t_m)$, we minimize cross-entropy against the predicted similarity distribution:
\begin{equation}
    \mathcal{L}_{temp\_v2a} = -\sum_{k=1}^{K} p_k
    \log \frac{\exp(S_{AV}(v, a_k) / \tau)}
    {\sum_{m=1}^{K} \exp(S_{AV}(v, a_m) / \tau)}.
\end{equation}
$\mathcal{L}_{temp\_a2v}$ uses audio clips as anchors, and $\mathcal{L}_{temp} = \frac{1}{2}(\mathcal{L}_{temp\_v2a} + \mathcal{L}_{temp\_a2v})$. This graduated supervision shapes a smooth alignment landscape in which similarity peaks at zero offset and decreases with displacement, mirrored by the empirical sensitivity curve in Fig.~\ref{fig:sensitivity}.

\section{Experimental Settings}

\subsection{Training Specification}

We initialize FATE from PE-AV-small~\citep{vyas2025pe-av}, discarding its text encoder and global pooling head, and fine-tune with LoRA~\citep{hu2022lora}, which updates only 2.3\% of the parameters while keeping the encoder weights frozen. Training uses the VGGSound~\citep{chen2020vggsound} training split (${\sim}$183k 10-second clips across 309 sound categories), where each video is segmented into 2-second clips with a 0.5-second stride, yielding ${\sim}$17 clips per video. The model is trained for 5 epochs on 8 NVIDIA A800 GPUs; preprocessing and optimization details are listed in Appendix.

\subsection{Evaluation Protocol}
We evaluate FATE on temporal cross-modal retrieval, zero-shot event localization, and generation evaluation, probing semantic-temporal discrimination, cross-task transfer, and agreement with human synchronization judgments, respectively.

\noindent\textbf{Temporal Cross-Modal Retrieval.}
Given an audio query $a_q$ and a video segmented into $N$ clips $\mathcal{V} = \{v_1, \dots, v_N\}$, the goal is to retrieve the temporally aligned clip:
\begin{equation}
    \hat{v} = \arg\max_{v_i \in \mathcal{V}} S_{AV}(v_i, a_q).
\end{equation}
We evaluate on the test splits of AVSync15~\citep{zhang2024avsync15} (150 videos) and
VGG-Sync~\citep{chen2021vggsync} (595 videos), each consisting of ${\sim}$10\,s videos. Both benchmarks are curated from VGGSound, so we verify by video ID that no test video appears in training split. Each video is segmented into 2s clips with a 0.5s stride, yielding $N{\approx}17$ candidates per video; one segment is selected as the ground-truth query, with graded relevance of 2 for the aligned segment, 1 for segments within $\pm$0.5\,s, and 0 otherwise. We report Recall@$k$ (R@$k$) and NDCG@$k$ (N@$k$) in both Video-to-Audio (V2A) and Audio-to-Video (A2V) directions.

We design two retrieval settings to disentangle model capabilities. In the \textbf{Intra-Video} setting, the candidate pool contains only the ${\sim}$17 segments of the same video; since all candidates share identical semantic content, this purely tests fine-grained temporal discrimination. In the \textbf{Inter-Video} setting, we append all segments from $M{=}50$ randomly sampled distractor videos (${\sim}$850 candidates per query), requiring the model to jointly identify the correct video and locate the aligned segment within it. Together, the two settings reveal whether a model possesses one capability or both.

We compare against two categories of baselines. The first consists of audio-visual foundation models, including ImageBind~\citep{girdhar2023imagebind}, LanguageBind~\citep{zhu2023languagebind}, CAV-MAE Sync~\citep{araujo2025cav-mae-sync}, PE-AV~\citep{vyas2025pe-av}, CAVP~\citep{luo2023cavp}, and PEAVS~\citep{goncalves2024peavs}, scored by cosine similarity between query and candidate. The second is the specialized synchronization model Synchformer~\citep{iashin2024synchformer}, which predicts temporal offsets rather than embeddings; we use its predicted probability at zero offset as the retrieval score. All baselines use officially released checkpoints.

\noindent\textbf{Audio-Visual Event Localization.}
Given an audio query $A_e$ and an video $V$, we divide $V$ into 1s segments and slide a window of the query's duration over them, predicting the window with the highest average similarity:
\begin{equation}
    \hat{t} = \arg\max_{t} \frac{1}{|w|}\sum_{s \in w(t)}
    S_{AV}(V_s, A_e).
\end{equation}

We evaluate on the Audio-Visual Event (AVE) dataset~\citep{tian2018ave}, which contains unconstrained videos with second-level temporal annotations across 28 event categories, and report segment-level accuracy for A2V, V2A, and their average. We compare against DCCA~\citep{dcca}, AVDLN~\citep{tian2018ave}, and DAM~\citep{wu2019dam}. Notably, all three baselines are \emph{supervised}, trained on the AVE training set with ground-truth temporal annotations, whereas FATE is evaluated \emph{zero-shot}, trained only on VGGSound without any AVE supervision.

\noindent\textbf{Joint Audio-Video Generation Evaluation.}
We evaluate FATE as an automatic metric for audio-visual
synchronization by measuring agreement between its score
$S_{AV}(V_{\mathrm{gen}}, A_{\mathrm{gen}})$ and human judgments
using Spearman's rank correlation $\rho$.
Using text prompts from the AVSync-15 test set, we generate
150 videos per model (750 in total) with five joint audio-video
generation models: BridgeDiT~\citep{guan2025bridgedit},
JavisDiT~\citep{liu2025javisdit}, Ovi~\citep{low2025ovi},
LTX-2~\citep{hacohen2026ltx2}, and JointDiT~\citep{wang2025jointdit}.
Ten independent annotators rate each video's temporal
synchronization on a 1--5 scale; their average gives its mean
opinion score (MOS).
The annotation protocol is detailed in Appendix A.3.

We compare FATE against DeSync~\citep{iashin2024synchformer},
PEAVS~\citep{goncalves2024peavs},
AV-Align~\citep{yariv2023avalign}, and CAVP~\citep{luo2023cavp}
at two levels.
\emph{Sample-level} $\rho$ compares metric scores with human MOS
across the 150 videos from each generation model, testing whether
the metric ranks individual videos consistently with humans.
\emph{Model-ranking} $\rho$ first averages metric scores and MOS
separately over each model's videos, then correlates the five
pairs of averages to assess agreement on the overall model ranking.
\definecolor{gray}{RGB}{232,232,232}
\definecolor{blue}{RGB}{225,237,250}

\begin{table*}[t]
    \centering
    \setlength{\tabcolsep}{2pt}
    \renewcommand{\arraystretch}{1.05}
    \caption{Temporal audio-visual cross-modal retrieval results. Intra-Video: candidates are segments from the same video ($\sim$17 per query), testing pure temporal discrimination. Inter-Video: candidates additionally include segments from $M{=}50$ distractor videos ($\sim$850 per query), requiring joint semantic and temporal discrimination. We report Recall@$k$ (R@$k$) and NDCG@$k$ (N@$k$) for Video-to-Audio (V2A) and Audio-to-Video (A2V) tasks. Best in bold, second best underlined.}
    \resizebox{\textwidth}{!}{%
    \begin{tabular}{l cccc cccc cccc cccc}
        \toprule
        & \multicolumn{8}{c}{\textbf{Avsync15}} & \multicolumn{8}{c}{\textbf{VGG-Sync}} \\
        \cmidrule(lr){2-9} \cmidrule(lr){10-17}
        & \multicolumn{4}{c}{V2A} & \multicolumn{4}{c}{A2V} & \multicolumn{4}{c}{V2A} & \multicolumn{4}{c}{A2V} \\
        \cmidrule(lr){2-5} \cmidrule(lr){6-9} \cmidrule(lr){10-13} \cmidrule(lr){14-17}
        \textbf{Method}
        & R@1 & R@3 & N@1 & N@3
        & R@1 & R@3 & N@1 & N@3
        & R@1 & R@3 & N@1 & N@3
        & R@1 & R@3 & N@1 & N@3 \\
        \midrule

        \rowcolor{gray}
        \multicolumn{17}{l}{\emph{Intra-Video} ($\sim$17 candidates per query)} \\
        ImageBind~\citep{girdhar2023imagebind}
            & 4.89 & 16.22 & 8.89 & 15.49 & 6.67 & 16.89 & 12.00 & 15.47
            & 6.19 & 19.19 & 10.31 & 16.50 & 7.25 & 20.67 & 10.70 & 17.35 \\
        LanguageBind~\citep{zhu2023languagebind}
            & 5.11 & 18.89 & 7.78 & 14.81 & 0.00 & 4.22 & 0.00 & 2.81
            & 5.43 & 17.09 & 8.91 & 14.48 & 0.06 & 3.59 & 0.06 & 2.15 \\
        CAV-MAE Sync~\citep{araujo2025cav-mae-sync}
            & 5.56 & 18.22 & 9.56 & 14.87 & 4.00 & 14.00 & 6.22 & 11.62
            & 5.15 & 15.29 & 9.30 & 13.42 & 4.15 & 14.68 & 7.96 & 12.35 \\
        PE-AV~\citep{vyas2025pe-av}
            & 5.33 & 18.00 & 9.33 & 15.56 & 6.22 & 17.33 & 11.11 & 15.95
            & 7.17 & 20.28 & 11.88 & 17.56 & 5.77 & 17.93 & 9.02 & 15.23 \\
        CAVP~\citep{luo2023cavp}
            & 6.44 & 16.22 & 9.56 & 14.28 & 3.33 & 14.22 & 6.89 & 11.80
            & 4.99 & 15.97 & 7.45 & 12.82 & 4.76 & 15.97 & 7.79 & 12.95 \\
        PEAVS~\citep{goncalves2024peavs}
            & 12.45 & 28.57 & 15.14 & 19.47 & 11.76 & 27.57 & 13.63 & 18.23
            & 11.05 & 25.85 & 16.34 & 20.35 & 10.75 & 24.85 & 13.71 & 16.19 \\
        Synchformer~\citep{iashin2024synchformer}
            & \underline{15.56} & \underline{38.89} & \underline{28.44} & \underline{38.46}
            & \underline{13.78} & \underline{32.89} & \underline{29.78} & \underline{34.98}
            & \underline{12.04} & \underline{34.12} & \underline{22.35} & \underline{31.57}
            & \underline{10.92} & \underline{32.21} & \underline{21.46} & \underline{30.97} \\
        \midrule
        \rowcolor{blue}
        \textbf{FATE (Ours)}
            & \textbf{22.37} & \textbf{55.74} & \textbf{53.56} & \textbf{59.24}
            & \textbf{21.56} & \textbf{56.22} & \textbf{51.78} & \textbf{59.66}
            & \textbf{26.39} & \textbf{48.37} & \textbf{60.83} & \textbf{69.04}
            & \textbf{16.50} & \textbf{44.83} & \textbf{33.61} & \textbf{44.49} \\
         \quad \textit{w/o frame-level}
            & 9.78 & 29.11 & 18.67 & 27.73 & 9.11 & 29.33 & 18.00 & 25.75
            & 12.72 & 34.68 & 22.80 & 32.82 & 10.76 & 28.68 & 19.27 & 26.55 \\

       \midrule
        \rowcolor{gray}
        \multicolumn{17}{l}{\emph{Inter-Video} ($\sim$850 candidates per query, $M{=}50$)} \\
        ImageBind~\citep{girdhar2023imagebind}
            & 2.89 & 9.78 & 5.11 & 9.96 & 2.89 & 7.78 & 5.11 & 7.24
            & 2.80 & 7.73 & 4.37 & 6.71 & 3.03 & 7.90 & 5.71 & 7.56 \\
        LanguageBind~\citep{zhu2023languagebind}
            & 2.35 & 4.80 & 6.15 & 8.90 & 1.95 & 4.10 & 5.60 & 7.45
            & 3.10 & 5.65 & 7.80 & 10.20 & 2.75 & 4.90 & 6.45 & 8.80 \\
        CAV-MAE Sync~\citep{araujo2025cav-mae-sync}
            & 0.22 & 0.44 & 0.67 & 0.56 & 0.22 & 0.44 & 0.22 & 0.40
            & 0.06 & 0.11 & 0.06 & 0.07 & 0.06 & 0.39 & 0.17 & 0.31 \\
        PE-AV~\citep{vyas2025pe-av}
            & 3.33 & 10.54 & 5.56 & 9.05 & \underline{4.67} & 10.12 & \underline{8.22} & \underline{10.21}
            & 4.15 & 10.70 & 6.61 & 9.20 & 2.41 & 7.11 & 3.98 & 6.14 \\
        CAVP~\citep{luo2023cavp}
            & 3.15 & 9.42 & 5.05 & 6.85 & 2.68 & 8.35 & 4.77 & 5.92
            & 2.95 & 8.15 & 4.23 & 6.10 & 2.45 & 7.20 & 4.15 & 5.25 \\
        PEAVS~\citep{goncalves2024peavs}
            & 3.68 & 12.64 & 6.44 & 10.29 & 2.99 & \underline{10.34} & 5.75 & 9.23
            & \underline{5.05} & \underline{15.85} & \underline{7.98} & \underline{13.35}
            & \underline{4.75} & \underline{14.85} & \underline{8.74} & \underline{13.19} \\
        SyncFormer~\citep{iashin2024synchformer}
            & \underline{4.44} & \underline{13.56} & \underline{8.00} & \underline{11.98}
            & 2.44 & 3.56 & 6.00 & 5.17 & 4.82 & 12.49 & 7.96 & 10.78
            & 0.90 & 2.07 & 1.90 & 2.58 \\
        \midrule
        \rowcolor{blue}
        \textbf{FATE (Ours)}
            & \textbf{14.22} & \textbf{34.13} & \textbf{33.53} & \textbf{37.66}
            & \textbf{11.11} & \textbf{30.44} & \textbf{23.11} & \textbf{30.11}
            & \textbf{15.35} & \textbf{37.76} & \textbf{35.85} & \textbf{42.24}
            & \textbf{11.93} & \textbf{31.54} & \textbf{24.93} & \textbf{31.99} \\
         \quad \textit{w/o frame-level}
            & 6.67 & 19.33 & 12.44 & 18.83 & 6.44 & 20.44 & 12.22 & 17.54
            & 7.90 & 22.02 & 14.06 & 20.87 & 7.23 & 19.44 & 13.28 & 18.34 \\
        \bottomrule
    \end{tabular}%
    }

    \label{tab:main_retrieval}
\end{table*}

\section{Experimental Results}


\subsection{Temporal Cross-Modal Retrieval}

Table~\ref{tab:main_retrieval} provides a controlled diagnosis of audio--visual retrieval: intra-video retrieval isolates \emph{when} an event occurs, whereas inter-video retrieval requires identifying both \emph{what} occurs and \emph{when}.

\noindent\textbf{Global pooling provides little usable temporal information.}
In the intra-video setting, all candidates share the same video-level semantics, and chance performance is 5.9\% R@1. Global-embedding baselines remain near this level on AVSync-15 (e.g., 4.89 for ImageBind and 5.33 for PE-AV), whereas FATE reaches 22.37 R@1 and 55.74 R@3. The same trend holds on VGG-Sync and for A2V retrieval, showing that semantic similarity alone cannot reliably localize events in time.

\noindent\textbf{Temporal supervision and frame-level structure provide complementary gains.}
With the same PE-AV backbone, applying our objectives without frame-level tokens raises intra-video V2A R@1 from 5.33 to 9.78 on AVSync-15. Restoring frame-level tokens further increases it to 22.37. The corresponding progression on VGG-Sync is similarly consistent: $7.17 \rightarrow 12.72 \rightarrow 26.39$. This controlled comparison separates the two effects: temporal supervision introduces synchronization awareness, while frame-level alignment preserves the temporal resolution needed to exploit it.

\noindent\textbf{Realistic retrieval requires both semantic grounding and temporal precision.}
With approximately 850 candidates, FATE achieves 34.13 V2A R@3 on AVSync-15 and 37.76 on VGG-Sync. The strongest baselines reach only 13.56 and 15.85, respectively, and the same ordering holds for A2V. For example, Synchformer decreases from 38.89 R@3 intra-video to 13.56 inter-video on AVSync-15 once semantic distractors are introduced. Thus, recognizing temporal offsets within a known video is insufficient: a transferable audio--visual embedding must represent both \emph{what} event occurs and \emph{when} it occurs.
\definecolor{gray}{RGB}{232,232,232}
\definecolor{blue}{RGB}{225,237,250}

\makeatletter
\newcommand{\continuousrowcolor}[1]{%
    \smash{\rlap{%
        \color{#1}%
        \rule[-\dp\@arstrutbox]
             {\linewidth}
             {\dimexpr\ht\@arstrutbox+\dp\@arstrutbox\relax}%
    }}%
}
\makeatother

\begin{table}[t]
    \centering
    \begin{minipage}[t]{0.44\linewidth}
        \vspace{0pt}
        \centering

        \caption{
        Audio-visual event localization on AVE.
        The upper and lower blocks contain supervised and
        zero-shot methods, respectively.
        Best and second-best results are bold and underlined.
        }
        \label{tab:matching}

        \footnotesize
        \setlength{\tabcolsep}{3pt}
        \renewcommand{\arraystretch}{1.08}

        \begin{tabular*}{\linewidth}{
            @{\extracolsep{\fill}} lccc
        }
            \toprule
            \textbf{Method}
            & \textbf{A2V}
            & \textbf{V2A}
            & \textbf{Avg.}\\
            \midrule

            \rowcolor{gray}
            \multicolumn{4}{l}{
                \emph{Supervised on AVE}
            }\\

            DCCA~\citeyearpar{dcca}
            & 34.8 & 34.1 & 34.5\\

            AVDLN~\citeyearpar{tian2018ave}
            & 44.8 & 35.6 & 40.2\\

            DAM~\citeyearpar{wu2019dam}
            & \underline{48.5}
            & \textbf{47.1}
            & \underline{47.8}\\

            \midrule
            \rowcolor{gray}
            \multicolumn{4}{l}{
                \emph{Zero-shot on VGGSound}
            }\\

            ImageBind~\citeyearpar{girdhar2023imagebind}
            & 23.4 & 19.8 & 21.6\\

            Synchformer~\citeyearpar{iashin2024synchformer}
            & 28.5 & 25.1 & 26.8\\

            PE-AV~\citeyearpar{vyas2025pe-av}
            & 34.2 & 29.4 & 31.8\\

            \cmidrule(lr){1-4}

            \continuousrowcolor{blue}
            \textbf{FATE (Ours)}
            & \textbf{51.8}
            & \underline{44.7}
            & \textbf{48.3}\\

            \quad \textit{w/o frame-level}
            & 37.9 & 33.2 & 35.6\\

            \bottomrule
        \end{tabular*}
    \end{minipage}
    \hfill
    \begin{minipage}[t]{0.53\linewidth}
        \vspace{0pt}
        \centering

        \caption{
        Ablation of FATE's core design choices on AVSync-15
        inter-video retrieval. Parentheses show relative drops
        from the full model.
        }
        \label{tab:ablation_design}

        \footnotesize
        \setlength{\tabcolsep}{3pt}
        \renewcommand{\arraystretch}{1.08}

        \begin{tabular*}{\linewidth}{
            @{\extracolsep{\fill}} lcc
        }
            \toprule
            \textbf{Variant}
            & \textbf{V2A R@3}
            & \textbf{A2V R@3}\\
            \midrule

            \continuousrowcolor{blue}
            \textbf{FATE (full)}
            & \textbf{34.13}
            & \textbf{30.44}\\

            \midrule
            \rowcolor{gray}
            \multicolumn{3}{l}{
                \emph{(i) Temporal label design}
            }\\

            \quad Hard binary (0/1)
            & 28.50 {\color{black!55}\scriptsize($-16.5\%$)}
            & 24.35 {\color{black!55}\scriptsize($-20.0\%$)}\\

            \quad Soft label, 0.25\,s
            & 29.78 {\color{black!55}\scriptsize($-12.7\%$)}
            & 28.12 {\color{black!55}\scriptsize($-7.6\%$)}\\

            \quad Soft label, 0.75\,s
            & 31.56 {\color{black!55}\scriptsize($-7.5\%$)}
            & 28.67 {\color{black!55}\scriptsize($-5.8\%$)}\\

            \midrule
            \rowcolor{gray}
            \multicolumn{3}{l}{
                \emph{(ii) Alignment strategy}
            }\\

            \quad Linear interpolation
            & 30.12 {\color{black!55}\scriptsize($-11.7\%$)}
            & 26.55 {\color{black!55}\scriptsize($-12.8\%$)}\\

            \quad Repeat padding
            & 29.40 {\color{black!55}\scriptsize($-13.9\%$)}
            & 25.80 {\color{black!55}\scriptsize($-15.2\%$)}\\

            \quad Audio$\rightarrow$video
            & 29.74 {\color{black!55}\scriptsize($-12.9\%$)}
            & 26.13 {\color{black!55}\scriptsize($-14.2\%$)}\\

            \midrule
            \rowcolor{gray}
            \multicolumn{3}{l}{
                \emph{(iii) Loss components}
            }\\

            \quad $\mathcal{L}_{\mathrm{sem}}$ only
            & 19.78 {\color{black!55}\scriptsize($-42.0\%$)}
            & 11.46 {\color{black!55}\scriptsize($-62.4\%$)}\\

            \quad $\mathcal{L}_{\mathrm{temp}}$ only
            & 15.56 {\color{black!55}\scriptsize($-54.4\%$)}
            & 12.34 {\color{black!55}\scriptsize($-59.5\%$)}\\

            \bottomrule
        \end{tabular*}
    \end{minipage}
    \vspace{-5mm}
\end{table}

\subsection{Cross-Modal Event Localization}
Table~\ref{tab:matching} evaluates zero-shot event localization on AVE and yields two findings.
First, among zero-shot methods, temporal structure in the representation is the decisive factor. All four share VGGSound training data and see no AVE annotations, so their ordering isolates representation design: ImageBind's global embeddings barely ground events in time (21.6\%), Synchformer's offset prediction helps (26.8\%), and FATE, which differs from PE-AV (31.8\%) only in preserving frame-level structure, reaches 48.3\%, a 16.5-point gain with an identical encoder.
Second, the learned temporal correspondence transfers beyond its training domain. Without training on AVE, FATE achieves 48.3\% average accuracy under our audio-query localization protocol, outperforming the evaluated zero-shot baselines. Compared with PE-AV using the same backbone, preserving frame-level temporal structure improves the average accuracy from 31.8\% to 48.3\%, demonstrating effective transfer of the learned temporal correspondence.

\subsection{Joint Audio-Video Generation Evaluation}

\definecolor{gray}{RGB}{232,232,232}
\definecolor{blue}{RGB}{225,237,250}

\begin{wraptable}{r}{0.49\textwidth}
    \vspace{-0.6\baselineskip}
    \centering

    \captionsetup{
        font=footnotesize,
        skip=3pt
    }

    \caption{
        Agreement with prompt-conditional human synchronization
        rankings on AVSync-15. Results are reported as average
        Spearman correlation $\bar{\rho}\times100$.
        Best and second-best results are bold and underlined.
    }
    \label{tab:generation_eval}

    \footnotesize
    \setlength{\tabcolsep}{3pt}
    \renewcommand{\arraystretch}{1.10}

    \begin{tabularx}{\linewidth}{
        @{}
        l
        >{\centering\arraybackslash}X
        >{\centering\arraybackslash}X
        @{}
    }
        \toprule

        \textbf{Metric}
        & \shortstack{
            \textbf{Cross-prompt}\\
            \textbf{relative-rank}\\
            $\boldsymbol{\bar{\rho}\times100\uparrow}$
        }
        & \shortstack{
            \textbf{Within-prompt}\\
            \textbf{model-rank}\\
            $\boldsymbol{\bar{\rho}\times100\uparrow}$
        }\\

        \midrule

        DeSync~\citeyearpar{iashin2024synchformer}
        & 8.22
        & 24.13\\

        PEAVS~\citeyearpar{goncalves2024peavs}
        & 8.04
        & 15.71\\

        AV-Align~\citeyearpar{yariv2023avalign}
        & 8.04
        & \underline{24.99}\\

        CAVP~\citeyearpar{luo2023cavp}
        & \underline{14.15}
        & 14.21\\

        \midrule

        \rowcolor{blue}
        \textbf{FATE (Ours)}
        & \textbf{17.24}
        & \textbf{44.41}\\

        \bottomrule
    \end{tabularx}
\end{wraptable}
Table~\ref{tab:generation_eval} evaluates agreement with human synchronization rankings. Since annotators rank five outputs within each prompt, we report two averaged Spearman correlations: \emph{cross-prompt relative-rank} correlation, computed across prompts for each generator and then averaged over generators, and \emph{within-prompt model-rank} correlation, computed among the fiveoutputs for each prompt and then averaged over prompts.
FATE performs best on both measures, achieving 17.24 in cross-prompt correlation and 44.41 in within-prompt correlation, compared with 14.15 for CAVP and 24.99 for AV-Align, respectively. This suggests that FATE more closely reproduces human relative rankings, particularly when comparing outputs generated from the same prompt. Nevertheless, the absolute correlations remain moderate, and each within-prompt correlation contains only five videos. We therefore interpret these results as evidence of relative ranking agreement rather than calibrated prediction of absolute synchronization quality.

\subsection{Ablation Study}
\noindent\textbf{Impact of Frame-Level Embeddings.}
To disentangle training from representation design, we compare FATE with the \emph{w/o frame-level} variant in Table~\ref{tab:main_retrieval}, trained using identical data, backbone, LoRA configuration, and budget. Replacing frame-level tokens with global pooling consistently reduced V2A R@3: from 55.74 to 29.11 intra-video and from 34.13 to 19.33 inter-video on AVSync-15, with the same trend on VGG-Sync (48.37 to 34.68 and 37.76 to 22.02). The pooled variant nevertheless slightly surpassed Synchformer on VGG-Sync intra-video R@1 (12.72 vs.\ 12.04), showing that the joint objective alone learns useful synchronization cues. Restoring frame-level structure raised R@1 to 26.39. Thus, the objective learns synchronization cues, while frame-level alignment provides the temporal precision needed for accurate retrieval.

\begin{minipage}[t]{0.48\linewidth}
\vspace{0pt}
\textbf{Temporal Sensitivity Analysis.}
We plot $S_{AV}$ against temporal offset $\Delta t$, averaged over all 150 AVSync-15 test videos. PE-AV with global pooling is nearly offset-invariant: even a 4-second shift yields almost the same score as perfect alignment. FATE instead exhibits a clear maximum at $\Delta t=0$, followed by a steady decay in both directions. This behavior follows directly from frame-level matching (Eq.~\ref{eq:frame_level}): shifting either modality breaks correspondence between physical moments and lowers the mean similarity. Its gradual decay further reflects the offset-aware soft labels in Eq.~\ref{eq:soft_label}, suggesting that FATE captures decreasing temporal overlap rather than imposing a binary aligned--misaligned boundary.
\end{minipage}\hfill%
\begin{minipage}[t]{0.5\linewidth}
\vspace{0pt}
\centering
\includegraphics[width=\linewidth]{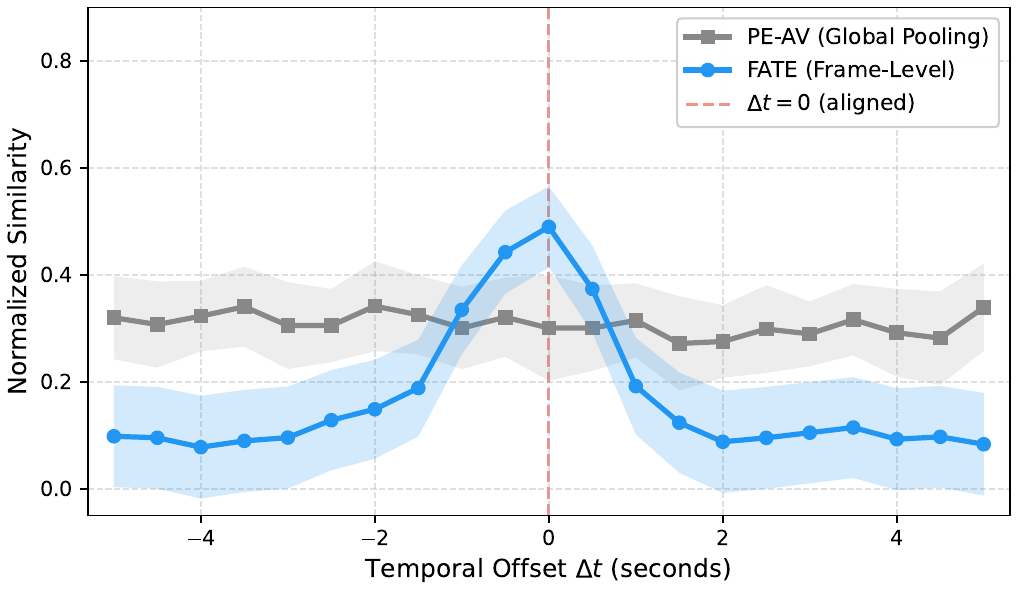}
\captionof{figure}{Temporal sensitivity of normalized audio--visual similarity. Shading denotes $\pm1$ standard deviation.}
\label{fig:sensitivity}
\end{minipage}


\noindent\textbf{Effectiveness of Loss Components.}
Table~\ref{tab:ablation_design}~(iii) evaluates each loss term.
Training with $\mathcal{L}_{sem}$ alone yields 19.78 V2A R@3: the model
learns cross-video semantics but lacks within-video temporal precision.
Training with $\mathcal{L}_{temp}$ alone performs worse still (15.56 V2A
R@3), as it provides no semantic signal to separate different videos.
Combining both reaches 34.13 V2A R@3, exceeding the sum of their individual
gains over PE-AV. This super-additive effect indicates the two objectives are
complementary: $\mathcal{L}_{sem}$ structures the embedding space across
videos, while $\mathcal{L}_{temp}$ refines temporal ordering within each
video.

\noindent\textbf{Soft vs.\ Hard Temporal Labels.}
Table~\ref{tab:ablation_design} (i) compares our raised-cosine soft targets,
which yield relative weights of $2{:}1{:}0$ at $|\Delta t|=0$, $0.5$\,s, and
$\geq1$\,s, against hard labels (aligned=1, else=0).
Hard labels degrade V2A R@3 from 34.13 to 28.50 and A2V R@3 from 30.44 to
24.35. The reason is straightforward: with a 0.5-second stride, adjacent
segments share 75\% temporal overlap with the anchor, and treating them as
hard negatives forces the model to push apart near-identical content,
introducing conflicting gradients. Soft labels instead assign intermediate
supervision to these ambiguous neighbors, letting the model learn a smooth
similarity decay rather than an abrupt boundary. Varying the soft-label
window confirms the choice: a narrower 0.25\,s window (29.78 V2A R@3)
mislabels positives as negatives, a wider 0.75\,s window (31.56) blurs
the boundary.

\noindent\textbf{Influence of Alignment Strategy.}
Table~\ref{tab:ablation_design} (ii) evaluates alternatives to
nearest-neighbor interpolation for aligning the higher-rate visual sequence
($T_v$ tokens) to the lower-rate audio sequence ($T_a$ tokens). Linear
interpolation (30.12 V2A R@3) averages adjacent frames, producing blended
features that correspond to no real frame and weaken temporal boundaries.
Repeat padding (29.40) duplicates visual tokens to match $T_a$, skewing the
similarity computation toward repeated content. Nearest-neighbor (34.13)
selects the closest original frame for each audio time step, so every
aligned token remains an unmodified representation. Reversing the resampling
direction (audio$\rightarrow$video, 29.74) also underperforms, as upsampling
audio degrades transient synchronization cues, supporting audio as the
proper temporal anchor.


\section{Conclusion}
We propose FATE, an audio-visual embedding model that unifies semantic correspondence and temporal synchronization within a single representation space. By preserving dense frame-level features, aligning them on the physical time axis, and training with a joint semantic-temporal objective, FATE avoids the information bottleneck imposed by global pooling. Experiments across three complementary tasks validate this design: FATE sets the state of the art on retrieval that demands both semantic and temporal discrimination, matches fully supervised methods on event localization in a purely zero-shot setting, and agrees most closely with human synchronization judgments as a generation evaluation metric. These results suggest that fine-grained temporal awareness can emerge from a general-purpose embedding space without task-specific prediction heads.

\section*{AI Use Statement}

Generative AI tools were used solely for language polishing and improving
the clarity of the manuscript. They were not used to generate experimental
results, conduct data analysis, or make scientific conclusions. All content
was reviewed and verified by the authors.

\section*{Reproducibility Statement}

We provide the complete model architecture, training objective, data
preprocessing procedure, hyperparameters, and evaluation protocols in the
main paper and supplementary material. Training details are provided in
Appendix A.2, and the human annotation protocol is described in Appendix
A.3. Anonymous source code is also provided with the submission.

\section*{Ethics Statement}

Our human evaluation involves annotators assessing the audio-visual
synchronization quality of generated videos. Annotators were informed of
the evaluation task and compensated for their participation. No sensitive
personal information was collected. The evaluated content was screened to
avoid harmful or inappropriate material.

\bibliography{main}
\bibliographystyle{iclr2027_conference}

\appendix
\clearpage
\section{Supplementary Material for FATE}

\begin{figure*}[b!]
    \centering \includegraphics[width=\textwidth]{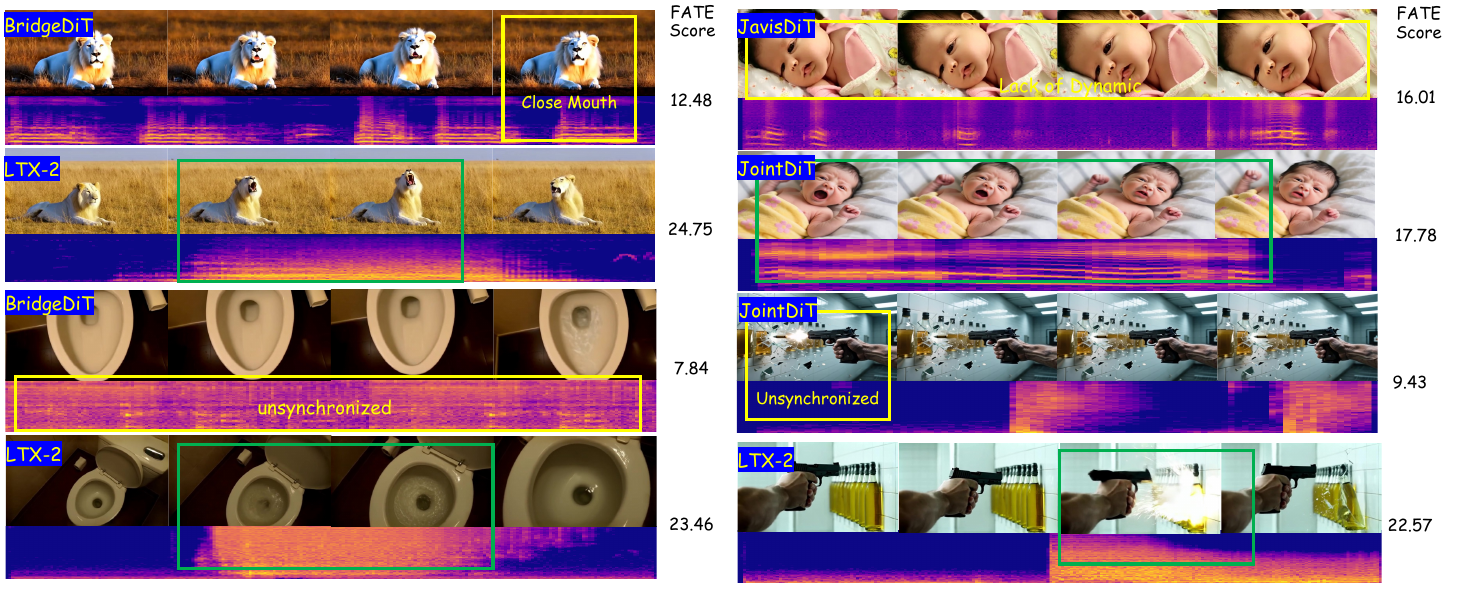}
    \caption{Qualitative visualization of FATE as a generation evaluation metric. For each case, we compare outputs from two generation models, showing key frames, mel-spectrograms, and FATE scores. FATE assigns higher scores to videos with well-aligned sound events (green boxes) and penalizes unsynchronized contents, closely matching human perceptual judgments.}
    \label{fig:case_study}
\end{figure*}

\subsection{Efficiency Analysis}
Table~\ref{tab:efficiency} compares the per-video inference time and embedding storage on AVSync-15. FATE (10.8\,s) is nearly identical to its backbone PE-AV (9.6\,s), as both share the dual-encoder architecture. In contrast, Synchformer requires 168.0\,s per video---roughly 15$\times$ slower---because its single-encoder architecture cannot pre-extract modality-independent embeddings, requiring a full forward pass for every query. This highlights a key advantage of FATE: it achieves superior temporal discrimination over Synchformer while maintaining better inference efficiency. Retaining frame-level embeddings increases per-video storage from 136\,KB to 3.3\,MB, since temporal information must be kept rather than pooled away. This remains an order of magnitude smaller than the raw video itself and is a justified trade-off for the fine-grained temporal capability that pooling-based models cannot provide, while query-time inference speed is preserved.
\begin{table}[h]
    \centering

    \footnotesize
    \setlength{\tabcolsep}{6pt}
    \renewcommand{\arraystretch}{1.05}

    \begin{tabular}{l cc}
        \toprule
        \textbf{Method} & \textbf{Storage (KB)}$\downarrow$ & \textbf{Time (s)}$\downarrow$ \\
        \midrule
        Global Pooling (PE-AV)                  & \textbf{136.0} & \textbf{9.6} \\
        SynchFormer~\citep{iashin2024synchformer} & --             & 168.0 \\
        \textbf{FATE (Ours)}                    & 3400.0         & 10.8 \\
        \bottomrule
    \end{tabular}
    \caption{Efficiency comparison on AVSync-15: average per-video inference time (in seconds) for feature extraction and cross-modal similarity computation, and per-video embedding storage ($\sim$10\,s videos, single NVIDIA A800 GPU). SyncFormer does not produce reusable embeddings, hence no storage entry.}
    \label{tab:efficiency}
\end{table}

\subsection{Training Details}
\label{app:training}
Table~\ref{tab:hyperparams} lists the complete preprocessing and optimization
configuration. Video frames are resized to $336 \times 336$ and audio is
resampled to 48\,kHz before feature extraction.

\begin{table}[t]
\centering
\small
\caption{Training hyperparameters of FATE.}
\label{tab:hyperparams}
\setlength{\tabcolsep}{5pt}
\renewcommand{\arraystretch}{1.05}
\begin{tabular}{ll}
\toprule
\textbf{Component} & \textbf{Setting} \\
\midrule
Backbone & PE-AV-small (visual + audio encoders) \\
Video resolution & $336 \times 336$ \\
Audio sampling rate & 48\,kHz \\
Clip segmentation & 2\,s window, 0.5\,s stride \\
\midrule
LoRA rank $r$ / $\alpha$ & 16 / 32 \\
LoRA dropout & 0.1 \\
LoRA target modules & Q, K, V, O projections (all attention layers) \\
Trainable parameters & 2.3\% \\
\midrule
Optimizer & AdamW ($\beta_1{=}0.9$, $\beta_2{=}0.999$) \\
Weight decay & 0.0 \\
Peak learning rate & $1 \times 10^{-4}$ \\
LR schedule & linear decay \\
Epochs & 5 \\
Precision & \texttt{bfloat16} \\
Per-GPU batch size & 4 \\
Gradient accumulation & 2 \\
GPUs & 8 $\times$ NVIDIA A800 \\
\midrule
Loss weights $\lambda_1$ / $\lambda_2$ & 0.5 (semantic) / 1.0 (temporal) \\
Temperature $\tau$ & 0.07 \\
\bottomrule
\end{tabular}
\end{table}

\subsection{Human Annotation Protocol}
\label{app:annotation}

\paragraph{Task setup.}
We collect human judgments on 750 generated videos, organized into 150 groups corresponding to the 150 text prompts from the AVSync-15 test set. Each group contains five videos, one from each generation model, allowing annotators to inspect outputs generated from the same prompt. Within each group, the videos are randomly shuffled and assigned filenames \texttt{1.mp4}--\texttt{5.mp4}. Model identities are hidden, and file indices do not consistently correspond to particular models across groups, preventing annotators from identifying generators through presentation order.

\paragraph{Evaluation criterion.}
Annotators assess \emph{audio-visual synchronization}, considering both temporal alignment and semantic consistency between sounds and visible events. Temporal alignment concerns whether a sound occurs at the same time as its corresponding visual event, without perceptible lead or lag. Semantic consistency concerns whether the sound is physically plausible for the visible event. For example, a barking sound should coincide with a dog's visible barking action; the presence of a dog alone does not establish temporal synchronization. Annotators are instructed to focus on these audio-visual relationships and disregard general content quality, visual aesthetics, and audio fidelity.

\paragraph{Annotation procedure.}
For each prompt, annotators view the five generated videos and assign a separate synchronization rating to each video on a 1--5 scale, with higher scores indicating better synchronization. Each video is independently evaluated by ten annotators. Let $r_{p,m}^{(k)} \in \{1,2,3,4,5\}$ denote the rating assigned by annotator $k$ to the video generated by model $m$ for prompt $p$, where $p \in \{1,\ldots,150\}$ and $m \in \{1,\ldots,5\}$. The mean opinion score (MOS) for this video is
\begin{equation}
\mathrm{MOS}_{p,m} = \frac{1}{10}\sum_{k=1}^{10} r_{p,m}^{(k)}.
\end{equation}
Thus, each of the 750 videos receives one MOS, obtained by averaging its ten individual ratings. This video-level MOS serves as the human reference score in our correlation analysis. Annotators are encouraged to replay videos and scrub the timeline around corresponding sound and visual events when their alignment is difficult to judge.

\subsection{Case Study}
Fig.~\ref{fig:case_study} provides qualitative evidence that FATE scores align with human perception of synchronization quality. We present four pairs of generated videos, each produced by two different generation models from the same text prompt. In the first example, BridgeDiT generates a lion video where the lion's mouth closes prematurely before the sound ends, receiving a low score of 12.48, while LTX-2 produces a version with well-aligned mouth movements and audio onsets, scoring 24.75. For the toilet flushing prompt (bottom-left), BridgeDiT generates audio that is clearly unsynchronized with the visual action (7.84), whereas LTX-2 produces a temporally coherent flush sequence (23.46). Beyond temporal misalignment, FATE also penalizes static visuals lacking dynamics (JavisDiT, 16.01) and missing sound-source correspondence (JointDiT, 9.43). Across all cases, the score rankings closely match human perceptual judgments, corroborating the quantitative results in Table~\ref{tab:generation_eval}.

\subsection{Prompts for MLLM-based Evaluation}
\label{app:mllm_prompt}

We benchmark FATE against two general-purpose omni-modal LLMs:
\texttt{Qwen3-Omni-30B-A3B} (open-weight, served locally with vLLM) and
\texttt{Gemini-3.6-Flash} (accessed through its official API). Both models
receive the identical zero-shot prompt shown below; no model-specific
tuning, few-shot exemplars, or chain-of-thought elicitation is used. The
prompt isolates temporal synchronization from semantic relevance, forbids
free-form explanation so that outputs can be parsed automatically, and fixes
the output to a two-decimal scalar in $[0,1]$ that is directly comparable to
normalized FATE scores.

Each video is submitted as a single request with its native audio track and
no additional text context. Decoding is greedy (temperature $0$, top-$p$
$1.0$, maximum 8 output tokens) to eliminate sampling variance, and each
video is queried once. Responses that fail to match the expected numeric
format are re-queried once; the small number of persistent failures are
excluded from the correlation analysis. Both models ingest the same
preprocessed clips used for FATE, so no model receives an input advantage.
We note that Qwen3-Omni samples video at 2\,fps, which upper-bounds the
temporal granularity it can in principle resolve; this sampling rate is a
property of the released model and cannot be increased through prompting.

\begin{promptbox}[title={Evaluation prompt for Omni LLM}]
\ttfamily\small\raggedright
You are an expert evaluator of audio-visual temporal synchronization.
\medskip

You will be given a short video clip with its audio track. Your task is to
rate ONLY the temporal synchronization between the audio and the visual
content: whether sound events occur at exactly the same moments as their
visible sources (e.g., a bark occurs while the dog's mouth opens, a drum hit
occurs at the moment of impact).
\medskip

\textbf{\rmfamily Scoring rules:}
\begin{itemize}
\setlength{\itemsep}{1pt}
\setlength{\parskip}{0pt}
\setlength{\topsep}{2pt}
  \item Output a single score between 0.00 and 1.00, with exactly two decimal places.
  \item 1.00 = perfectly synchronized: every audible event coincides with its visible cause.
  \item 0.50 = partially synchronized: some events align, others are noticeably shifted.
  \item 0.00 = completely unsynchronized: sound events and visible actions do not co-occur at all.
  \item Judge synchronization continuously; use the full range (e.g., 0.15, 0.38, 0.85), not only the anchor values above.
\end{itemize}
\medskip

\textbf{\rmfamily Strict constraints:}
\begin{itemize}
\setlength{\itemsep}{1pt}
\setlength{\parskip}{0pt}
\setlength{\topsep}{2pt}
  \item Ignore semantic relevance. Even if the sound type matches the scene (a dog video with barking audio), the score must be LOW if the timing is off (e.g., barking while the mouth is closed).
  \item Ignore video quality, aesthetics, audio fidelity, and content plausibility.
  \item If the sounding source is never visible on screen, output 0.00.
  \item Do not explain. Output ONLY the score in the format: X.XX
\end{itemize}
\end{promptbox}

\end{document}